\documentclass[preprint,12pt]{elsarticle}

\usepackage{amssymb}

\journal{arXiv}

\begin{document}

\begin{frontmatter}



\title{Superstatistics with cut-off tails for financial time series}


\author[label1,label2]{Yusuke Uchiyama}
\author[label1]{Takanori Kadoya}

\address[label1]{MAZIN Inc., Toshima-ku, 170-0005 Tokyo, Japan}
\address[label2]{Faculty of Engineering, Information and Systems, University of Tsukuba, Tsukuba, Ibaraki, 305-8573 Japan}

\begin{abstract}
Financial time series have been investigated to follow fat-tailed distributions.  Further, an empirical probability distribution sometimes shows cut-off shapes on its tails.  To describe this stylized fact, we incorporate the cut-off effect in superstatistics.  Then we confirm that the presented stochastic model is capable of describing the statistical properties of real financial time series.  In addition, we present an option pricing formula with respect to superstatistics.
\end{abstract}


\end{frontmatter}


\section{Introduction}
\label{sec:sec1}
In financial markets, such as stocks, bonds, commodities and foreign exchanges,  prices fluctuate randomly.  Capturing the nature of the fluctuation of financial time series has been of interest for both industrial and private investors to design trading strategies.  Thus many have been attempted to accurately model the financial time series. \par
Brownian motion has been introduced as a stochastic model in finance because financial time series were also believed to folow the Gaussian statistical law \cite{Bachelier, Samuelson}.  However, actual financial time series have shown fat-tails in its probability density functions (PDFs) \cite{Mandelbrot}.  \par
Fat-tailed PDFs, which are considered to be anomalous fluctuations, are often observed in complex systems.  Such anomaly results from many body interaction of participants in the complex systems.  Financial markets can also be considered to be a many body interaction of traders.  Indeed, asset prices fluctuate, affected by trading strategies, traders' sentiment, macroeconomic news and so on.  Financial time series also show anomalous fluctuations, known as stylized facts \cite{Cont}  \par
Based on this idea, superstatistics have been developed as a superposition of fast and slow random variables \cite{Beck-Cohen}.  Dynamical foundations of nonextensive statistical mechanics can be formulated with superstatistics \cite{Beck1}.  Subsequently, anomalous fluctuations in various systems 
have been identified by the method of superstatistics \cite{Cohen1, Beck2}.  In the case of finance, volatility of return diverges as the result of superstatistics \cite{Straeten-Beck}.  Nevertheless, empirical volatility of return is estimated as a finite value.  \par
In this paper, we incorporate a cut-off effect in superstatistics to describe the stylized facts in financial markets.  The proposed stochastic model shows fat-tails with cut-off and then attempts to fit the fluctuations of a stock index.  In addition, we derive a formula for option pricing using the proposed stochastic model for the return of asset price.
\section{Stochastic models}
\label{sec:sec2}
\subsection{Superstatistics}
\label{sec:subsec2-1}
Fluctuations in complex systems consists of multiscale dynamics.  Based on this idea, Beck introduce superatatistics, which is the abbreviation for superposition of statistics \cite{Beck1}.  The key concept of superstatistics is local equilibrium, where a slow random variable is regarded as constant while a fast random variable fluctuates.  In this situation, a stochastic differential equation (SDE) presents the dynamics of the fast random variable whereas the slow random variable is given as a parameter of the SDE.   \par 
The Ornstein-Uhlenbeck(OU) process has been often used as the SDE  for superstatistics since its equilibrium distribution is a Gaussian distribution.  Statistical results from the Gaussian distribution are readily extended to non-Gaussian statistics.  The OU process for a random variable $X$ is a linear SDE represented as
\begin{equation}
dX=-{\gamma}Xdt+{\sigma}dW
\label{eq:OUP}
\end{equation}
with ${\gamma}$ and ${\sigma}$ being positive real parameters \cite{Uhlenbeck-Ornstein}.  The Fokker-Planck equation for the OU process in Eq. (\ref{eq:OUP}) gives the Gaussian distribution as the local equilibrium \cite{Riskin}:  
\begin{equation}
p(x|{\beta})=\sqrt{\frac{\beta}{\pi}}\;{\exp}(-{\beta}x^2),
\label{eq:GD}
\end{equation}
where ${\beta}$ is a slow random variable known as inverse temperature.  Note that Eq. (\ref{eq:GD}) is the conditional Gaussian distribution with respect to ${\beta}$.
When the PDF for ${\beta}$ is given as $f({\beta})$, the Bayes' theorem provides the marginal distribution for $X$ by 
\begin{equation}
p(x)=\int_0^{\infty}p(x|{\beta})f({\beta})d{\beta}.
\label{eq:SS}
\end{equation}
The specific form of $f({\beta})$ is identified by empirical data \cite{Straeten-Beck, Beck-Cohen-Swinney}.  The gamma, the inverse gamma and the log-normal distribution are often used as PDFs for $f({\beta})$, which are classified by log-amplitude cumulants \cite{Kiyono-Konno}.  \par
For financial time series, the slow fluctuations of the inverse temperature have been identified with the gamma distribution \cite{Straeten-Beck, Beck-Cohen-Swinney, Xu-Beck, Takaishi}.  Inserting the conditional Gaussian distribution in Eq. (\ref{eq:GD}) and the gamma distribution
\begin{equation}
f({\beta})=\frac{b^a}{{\Gamma}(a)}{\beta}^{a-1}{\exp}(-b{\beta})
\label{eq:Gamma}
\end{equation}
into Eq. (\ref{eq:SS}), one obtains the marginal PDF as 
\begin{equation}
p(x)=\frac{b^a}{\sqrt{2{\pi}}}\frac{{\Gamma}(a+\frac{1}{2})}{{\Gamma}(a)}\frac{1}{(b+x^2)^{a+\frac{1}{2}}},
\label{eq:SSTqD}
\end{equation}
where $a$ and $b$ are positive real parameters, and ${\Gamma}({\cdot})$ is the gamma function \cite{NIST}.  Note that the variance of the PDF in Eq. (\ref{eq:SSTqD}) is a finite value if$a{\geq}1$.  But for empirical data, the case that $a<1$ is sometimes observed.
\subsection{Cut-off effect}
\label{sec:subsec2-2}
To resolve the problem of infinite variance, we introduce the cut-off effect to superstatistics.  Supposing the inverse temperature is decomposed into a constant and a random part, the conditional Gaussian distribution in Eq. (\ref{eq:GD}) becomes
\begin{equation}
p(x|{\beta})=\sqrt{\frac{ {\beta}_0 + {\beta}}{\pi}}\;{\exp}\left[-({\beta}_0 + {\beta})x^2\right]
\label{eq:mGD}
\end{equation}
with ${\beta}_0$ and ${\beta}$ being respective constant and random parameters.  Implementing superposition with respect to ${\beta}$ gives
\begin{equation}
p(x)=\frac{1}{\sqrt{\pi}}{\exp}\left[-{\beta}_0x^2\right]\int_0^{\infty}\sqrt{{\beta}_0+{\beta}}\;{\exp}\left[-{\beta}x^2\right]f({\beta})d{\beta}. 
\label{eq:mSS}
\end{equation}
Here one can see that the Gaussian term with ${\beta}_0$ holds the variance of the PDF in Eq. (\ref{eq:mSS}) to be finite, even if the integrand has an algebraic order term.  The constant parameter ${\beta}_0$ indicates the magnitude of the cut-off effect since ${\beta}_0{\to}0$ leads to the ordinary superstatistics in Eq. (\ref{eq:SS}) .  \par
In the case of the random parameter ${\beta}$ following the gamma distributions in Eq. (\ref{eq:Gamma}), the marginal PDF in Eq. (\ref{eq:mSS}) is specifically obtained as
\begin{equation}
p(x)=\frac{b^a}{\sqrt{\pi}{\Gamma}(a)}{\Gamma}_g\left(a,-\frac{1}{2};{\beta}_0(x^2+b)\right)\frac{{\exp}(-{\beta}_0x^2)}{(x^2+b)^{a+\frac{1}{2}}}
\label{eq:TGCD}
\end{equation}
with a generalized gamma function defined by
\begin{equation}
{\Gamma}_g(z,{\lambda};v)=\int_0^{\infty}({\xi}+v)^{-{\lambda}}{\xi}^{z-1}{\rm e}^{-{\xi}}d{\xi},
\end{equation}
where ${\rm Re}(z)>0$ \cite{Kobayashi}.  In fact, the variance of the PDF in Eq. (\ref{eq:TGCD}) is evaluated as
\begin{equation}
{\langle}X^2{\rangle}=\frac{1}{2}\frac{b}{{\Gamma}(a)}{\Gamma}_g(a,-1;b{\beta}_0),
\label{eq:VTGCD}
\end{equation}
and then is a finite value (see \ref{sec:app1}).  Thus the presented model in Eq. (\ref{eq:TGCD}) is applicable to describe fluctuations in complex systems without the problem of infinite variance.  
\section{Empirical data analysis}
\label{sec:sec3}
To validate superstatistics with the cut-off effect, we implemented an empirical data analysis for a financial time series, S\&P 500.  As with previous works for financial time series, we used the log-return of the price data to eliminate trend components, and then estimated the empirical PDF.  \par
\begin{figure}
\begin{center}
\includegraphics[]{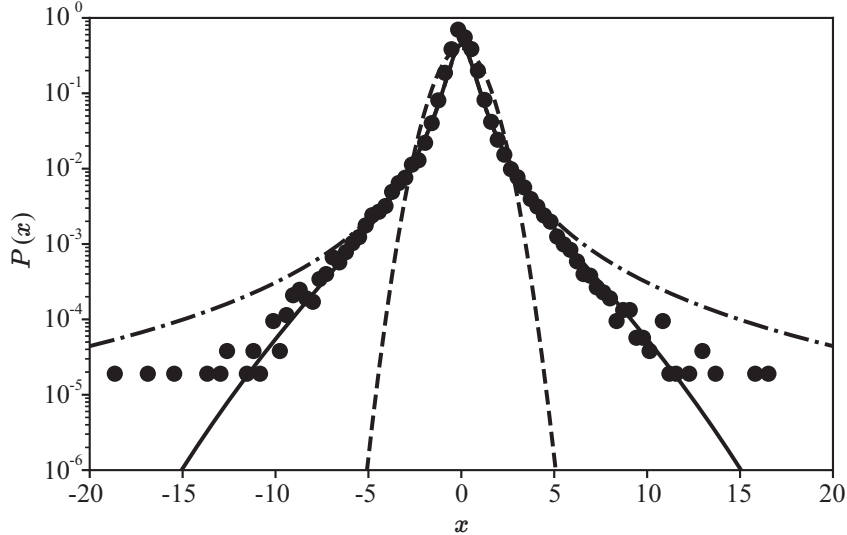}
\caption{\label{fig:PDF}PDF for log-return of S\&P500 from 1994 to 1999．The black circles indicate the empirical PDF. The solid line is the PDF in eq. (\ref{eq:TGCD}) with .$a=0.904$, $b=0.571$, ${\beta}_0=0.0252$  The dashed line is the Gaussian distribution.  The dashed-dotted line is the PDF in Eq. (\ref{eq:SSTqD}).  }
\end{center}
\end{figure}

In Fig. 1, the black circles are the empirical PDF, which is normalized by its empirical mean and variance, of S\&P 500 from 1992 to 1994 sampled each minute, the solid line is the PDF in Eq. (\ref{eq:TGCD}), the dashed line is the Gaussian distribution, and the dashed-dotted line is the PDF in Eq. (\ref{eq:SSTqD}).  The parameters of each PDF in Fig. \ref{fig:PDF} are estimated via the method of maximum likelihoods.  It was confirmed that the empirical PDF has fat-tails with smooth cut-off.  The PDF in Eq. (\ref{eq:SSTqD}) identifies well the shape of the empirical PDF on both the top and the tails, whereas others have been unable to closely capture the whole shape of the distribution.  Thus the proposed stochastic model captures the nature of fluctuations of the S\&P 500.  \par
In this case, we employed the gamma distribution as $f({\beta})$.  On the other hand, other PDFs are also applicable to describe fluctuations of financial time series.  In fact, the Ibovespa index, which is a Brazilian stock index, has been reported that empirical PDF of its log-return shows exponentially decaying tails with cut-off \cite{Sosa-Correa-Ramos-Vasconcelos}.  It is possible to employ the inverse gamma distribution as $f({\beta})$ in Eqs. (\ref{eq:SS}) or (\ref{eq:mSS}) for identifying the fluctuations.  

\section{Option pricing model}
\label{sec:sec4}
One of the useful applications of financial time series is option pricing, which was established by Black and Scholes \cite{Black-Scholes}.  In their work, geometric Brownian motion is employed as the asset price model.  
Then the price of a European call option, which is the right to buy an asset by strike price at the date of expiration, was derived as a closed form solution of the Black-Scholes equation.  However, as observed in the previous section, the empirical log-return process does not follow the Gaussian distribution. Here, we instead apply the present stochastic model to the pricing of European call options.  \par
Suppose the asset price is given by $S_t={\rm e}^{Y_t}S_0$, where $Y_t$ is the log-return of the asset price defined by $Y_t={\ln}S_{t+t_0}-{\ln}S_{t_0}$.  The evolution of $Y_t$ is assumed to be modelled by Brownian motion as
\begin{equation}
dY={\mu}dt+\frac{1}{\sqrt{2\beta}}dW,
\label{eq:SDELR}
\end{equation}
where ${\mu}$ is a constant parameter and ${\beta}$ is a slow random parameter.  Since ${\beta}$ fluctuates slower than $Y_t$, we can obtain a risk-free asset $P_t={\rm e}^{\left(r+\frac{1}{2{\beta}}\right)t}P_0$ with a risk-free rate $r$ while ${\beta}$ seems to be constant.  Then we construct a portfolio $V(Y_t,t)$ in $0{\leq}t{\leq}T$ as
\begin{equation}
V={\phi}S_t+{\psi}P_t,
\label{eq:PFL}
\end{equation}
where ${\phi}(t)$ and ${\psi}(t)$ are weight coefficients.  Since the components of the portfolio are assumed to be invariant for an infinitesimal time period, Eq. (\ref{eq:PFL}) satisfies the following differential form:
\begin{equation}
dV={\phi}dS_t+{\psi}dP_t.
\label{eq:DiffPFL}
\end{equation}
At the same time, It$\hat{\rm o}$'s formula for Eq. (\ref{eq:PFL}) provides
\begin{equation}
dV=\left[\left({\mu}-\frac{1}{2{\beta}}\right)\frac{{\partial}V}{{\partial}Y}+\frac{1}{2{\beta}}\frac{{\partial}^2V}{{\partial}Y^2}+\frac{{\partial}V}{{\partial}t}\right]dt+\frac{1}{\sqrt{\beta}}\frac{{\partial}V}{{\partial}Y}dW.
\label{eq:ItoV}
\end{equation}
Comparing Eqs. (\ref{eq:PFL}) with (\ref{eq:DiffPFL}) and setting
\begin{equation}
{\phi}=\frac{1}{S_0}{\rm e}^Y\frac{{\partial}V}{{\partial}Y}, 
\label{eq:phi}
\end{equation}
we obtain the following partial differential equation for the portfolio:
\begin{equation}
\frac{{\partial}V}{{\partial}t}+r\frac{{\partial}V}{{\partial}Y}
-\frac{1}{2{\beta}}\frac{{\partial}^2V}{{\partial}Y^2}-\left(r+\frac{1}{4{\beta}}\right)V=0,
\label{eq:ABFE}
\end{equation}
which is the Black-Scholes equation with respect to the log-return $Y_t$.  \par
The problem of option pricing is formalized as a boundary value problem for Eq. (\ref{eq:ABFE}).  Here we consider European call options with the following boundary conditions:
\begin{eqnarray}
&&\lim_{Y{\to}-{\infty}}V(Y,t)=0, \\
&&\lim_{Y{\to}{\infty}}V(Y,t)=S_0{\rm e}^{Y}, \\
&&V(Y_T,T)={\max}\{S_0{\rm e}^{Y_T}-K, 0\},
\end{eqnarray}
where $K$ is a strike price.  On the setup,  Eq. (\ref{eq:ABFE}) is solved as
\begin{equation}
V(Y,t)={\rm e}^{-r(T-t)}\int_{-{\infty}}^{\infty}{\rm e}^{-\frac{1}{4{\beta}}(T-t)}G(Y-Y_T,T-t)V(Y_T,T)dY_T
\label{eq:CVSOL}
\end{equation}
with $G(Y,t)$ being the Green function derived in \ref{sec:app2}.  This is a conditional stochastic variable with respect to ${\beta}$ and thus is rewritten as $V(Y,t|{\beta})$.  Given a PDF for ${\beta}$ as $f({\beta})$, the Bayes' theorem provides
\begin{eqnarray}
V(Y,t)
&=&\int_0^{\infty}V(Y,t|{\beta})f({\beta})d{\beta} \\
&=&{\rm e}^{-r(T-t)}\int_{-{\infty}}^{\infty}F(Y-Y_T,T-t)V(Y_T,T)dY_T,
\label{eq:VSOL}
\end{eqnarray}
where $F(Y,t)$ is the integral kernel defined by
\begin{equation}
F(Y,t)=\int_0^{\infty}{\rm e}^{-\frac{1}{4{\beta}}(T-t)}G(Y,T-t|{\beta})f({\beta})d{\beta}.
\label{eq:IKER}
\end{equation}
The closed form of Eq. (\ref{eq:IKER}) can be obtained when $f({\beta})$ is the gamma distribution or the inverse gamma distribution, which are presented in \ref{sec:app3}.  In addition, replacing ${\beta}$ in Eq. (\ref{eq:IKER}) with ${\beta}_0+{\beta}$, we can utilize the log-return model introduced in Eq. (\ref{eq:mSS}).
\section{Conclusion}
\label{sec:sec5} 
We incorporate the cut-off effect in superstatistics as a stochastic model which resolves the problem of infinite variance.  The analytical representation of the model was presented in the case of a random parameter following the gamma distribution.  It is confirmed that the Gaussian term holds the variance of the PDF to be finite, which is consistent with the empirical statistical results in real world.  In fact, the model was confirmed to closely identify well fluctuations of the S\&P 500. \par
In addition, the formula for European call option was derived from the Black-Scholoes equations with the proposed model for asset price, which is a natural extension of the ordinary option pricing model.  Closed forms can be obtained in the case of a random parameter following the gamma or the inverse gamma distribution while numerical integrals should be implemented with other PDFs.  The method presented can be applied exotic options with appropriate mathematical techniques, which will be our next challenge.   \par
As is well known, superstatistics have shown the power in various fields, in both natural and social science and in engineering.  Hence the presented model can also be applied to statistically analyse random phenomena in other systems.  

 \appendix

 \section{Variance of the PDF in Eq. (\ref{eq:TGCD})}
 \label{sec:app1}
Since the PDF in Eq. (\ref{eq:SS}) is symmetric, the variance is obtained as the second moment,
\begin{eqnarray}
{\langle}X^2{\rangle}
&=&\int_{-\infty}^{\infty}x^2p(x)dx \\
&=&\int_{-\infty}^{\infty}x^2\int_0^{\infty}p(x|{\beta})f({\beta})d{\beta}dx \nonumber \\
&=&\int_0^{\infty}\int_{-\infty}^{\infty}x^2p(x|{\beta})dxf({\beta})d{\beta} \nonumber \\
&=&\int_0^{\infty}{\langle}X^2{\rangle}_{\beta}f({\beta})d{\beta}, \nonumber
\end{eqnarray}
where ${\langle}X^2{\rangle}_{\beta}$ is the conditional variance of $X$ with respect to ${\beta}$.  The conditional variance of the Gaussian distribution in Eq. (\ref{eq:mGD}) is calculated as  
\begin{equation}
{\langle}X^2{\rangle}_{\beta}=\frac{1}{2}({\beta}_0+{\beta})^{-1},
\end{equation}
whereby the gamma distribution in Eq. (\ref{eq:Gamma}) with the conditional variance gives the variance of $p(x)$ in Eq. (\ref{eq:VTGCD}).  \par
To show that the variance in Eq. (\ref{eq:VTGCD}) is finite, we estimate the generalized gamma function, which is expressed by
\begin{equation}
{\Gamma}_g\left(a,-\frac{1}{2};{\beta}_0(x^2+b)\right)
=
\int_0^{\infty}({\eta}+b{\beta}_0)^{-1}{\eta}^{a-1}{\rm e}^{-{\eta}}d{\eta},
\end{equation}
where $x$ is transformed to $\sqrt{\eta}$.  When $a>1$, we employ the Schwartz inequality with respect to a probability measure ${\rm e}^{-{\eta}}$ as
\begin{equation}
\int_0^{\infty}({\eta}+b{\beta}_0)^{-1}{\eta}^{a-1}{\rm e}^{-{\eta}}d{\eta}
{\;\;}{\leq}{\;\;}
\left(\int_0^{\infty}({\eta}+b{\beta}_0)^{-2}{\rm e}^{-{\eta}}d{\eta}\right)^{\frac{1}{2}}
\left(\int_0^{\infty}{\eta}^{2(a-1)}{\rm e}^{-{\eta}}d{\eta}\right)^{\frac{1}{2}}.
\end{equation}
The integrals on the right hand side are respectively evaluated as
\begin{eqnarray}
\int_0^{\infty}({\eta}+b{\beta}_0)^{-2}{\rm e}^{-{\eta}}d{\eta}
&{\leq}&
\int_0^{\infty}({\eta}+b{\beta}_0)^{-2}d{\eta} \\
&=&
\frac{1}{3}(b{\beta}_0)^{-3} \nonumber
\end{eqnarray}
and
\begin{equation}
\int_0^{\infty}{\eta}^{2(a-1)}{\rm e}^{-{\eta}}d{\eta}={\Gamma}(2a-1).
\end{equation}
Hence, we obtain
\begin{equation}
\int_0^{\infty}({\eta}+b{\beta}_0)^{-1}{\eta}^{a-1}{\rm e}^{-{\eta}}d{\eta}<{\infty}.
\end{equation}
On the other hand, when $0<a<1$, the integral is evaluated as 
\begin{eqnarray}
\int_0^{\infty}({\eta}+b{\beta}_0)^{-1}{\eta}^{a-1}{\rm e}^{-{\eta}}d{\eta}
&{\leq}&
\int_0^{\infty}({\eta}+b{\beta}_0)^{-1}{\eta}^{a-1}d{\eta} \\
&=&
(b{\beta}_0)^{a-1}B(a,1-a), \nonumber
\end{eqnarray}
where $B({\cdot},{\cdot})$ is the beta function \cite{NIST}.
 \section{Green function of Eq. (\ref{eq:ABFE})}
 \label{sec:app2}
By change of variables,
 \begin{eqnarray}
 &&s=T-t, \\
 &&X=Y+rs, \\
 &&U(X,s)={\rm e}^{\left(r+\frac{1}{4{\beta}}\right)(T-t)}V(Y,t),
 \end{eqnarray}
 Eq. (\ref{eq:ABFE}) leads to the diffusion equation,
 \begin{equation}
 \frac{{\partial}U}{{\partial}s}=\frac{1}{4{\beta}}\frac{{\partial}^2U}{{\partial}X^2}.
\label{eq:DEQ} 
 \end{equation} 
 Hence the Green functions of Eq. (\ref{eq:DEQ}), $G(X,s)$, is given by the Gaussian function.  The specific form of the Green function of Eq. (\ref{eq:ABFE}) is readily obtained as
 \begin{equation}
 G(Y,T-t)=\sqrt{\frac{\beta}{{\pi}(T-t)}}{\exp}\left[-\frac{{\beta}(Y+r(T-t))^2}{T-t}\right].
 \label{eq:GFUNC}
 \end{equation} 
\section{Closed form of $F(Y,t)$ in Eq. (\ref{eq:IKER}) with the gamma or  the inverse gamma distribution}
\label{sec:app3}
The integral kernel $F(Y,t)$ in Eq. (\ref{eq:IKER}) with Eq. (\ref{eq:GFUNC}) is rewritten as
\begin{equation}
F(Y,t)=\int_0^{\infty}\sqrt{\frac{\beta}{{\pi}(T-t)}}\;{\exp}\left[-\frac{(Y+r(T-t))^2{\beta}}{T-t}-\frac{T-t}{4{\beta}}\right]f({\beta})d{\beta}.
\label{eq:FKER}
\end{equation}
To implement the integral in Eq. (\ref{eq:FKER}), we utilize an integral representation of the modified Bessel function as
\begin{equation}
K_{\nu}(z)=\frac{1}{2}\left(\frac{z}{2}\right)^{\nu}\int_0^{\infty}t^{-({\nu}+1)}{\exp}\left(-t-\frac{z^2}{4t}\right)dt
\label{eq:IMBSSL}
\end{equation}
with the condition that $|{\rm arg}z|<\frac{\pi}{4}$ \cite{NIST}.  In the case of ${\beta}$ following the gamma distribution,
\begin{equation}
f({\beta})=\frac{b^a}{{\Gamma}(a)}{\beta}^{a-1}{\exp}(-b{\beta}),
\end{equation}
Eq. (\ref{eq:FKER}) is obtained as
\begin{equation}
F(Y,t)=\frac{1}{{\Gamma}(a)}\sqrt{\frac{2}{\pi}}\left[\frac{b(T-t)}{2}\right]^a{\zeta}^{\frac{a}{2}}K_{-a-\frac{1}{2}}({\zeta}),
\end{equation}
where ${\zeta}$ is given by
\begin{equation}
{\zeta}(Y,t)=[Y+r(T-t)]^2+b(T-t).
\end{equation}
Likewise, when ${\beta}$ follows the inverse gamma distribution,
\begin{equation}
f({\beta})=\frac{b^a}{{\Gamma}(a)}{\beta}^{-a-1}{\exp}\left(-\frac{b}{\beta}\right),
\end{equation}
Eq. (\ref{eq:FKER}) is obtained as
\begin{equation}
F(Y,t)=\frac{(2b)^a}{{\Gamma}(a)}\sqrt{\frac{2}{{\pi}(T-t)}}\;{\xi}^{-\frac{a}{2}-\frac{1}{4}}K_{a-\frac{1}{2}}({\eta}),
\end{equation}
where ${\xi}$ and ${\eta}$ are respectively given by
\begin{eqnarray}
&&{\xi}(Y,t)=\frac{[Y+r(T-t)]^2}{(T-t)(T-t+b)}, \\
&&{\eta}(Y,t)=\sqrt{\frac{T-t+b}{T-t}}\;|Y+r(T-t)|.
\end{eqnarray}
%



\end{document}